\documentclass[pra,aps,twocolumn]{revtex4}
\usepackage{graphicx,amsmath}
\usepackage{bbm}
\usepackage{txfonts}
\usepackage{hyperref}

\newcommand{\rp}[1]{(\ref{#1})}

\newcommand{\abs}[1]{\left|{#1}\right|}

\newcommand{\av}[1]{\left\langle #1 \right\rangle}

\newcommand{\al}[1]{^{(#1)}}
\newcommand{\da}{^\dagger}

\newcommand{\pt}[1]{\left( #1 \right)}
\newcommand{\pq}[1]{\left[ #1 \right]}
\newcommand{\pg}[1]{\left\{ #1 \right\}}

\newcommand{\bs}[1]{\boldsymbol #1}

\newcommand{\lpg}[1]{\left\{ #1 \right.}

\newcommand{\rpg}[1]{\left. #1 \right\}}
\newcommand{\ee}{{\rm e}}
\newcommand{\ii}{{\rm i}}
\newcommand{\dd}{{\rm d}}

\newcommand{\nn}{{\nonumber}}

\newcommand{\mat}[2]{ 
                      \begin{array}{#1}
                       #2
                       \end{array}  }

\newcommand{\va}{{\bf a}}

\newcommand{\AAA}{{\cal A}}
\newcommand{\BB}{{\cal B}}
\newcommand{\CC}{{\cal C}}

\newcommand{\LL}{{\cal L}}
\newcommand{\MM}{{\cal M}}
\newcommand{\NN}{{\cal N}}

\newcommand{\VV}{{\cal V}}

\begin{document}

\title{Mechanical EPR entanglement with a finite-bandwidth squeezed reservoir}

\author{Muhammad~Asjad, Stefano~Zippilli, David~Vitali}
\affiliation{School of Science and Technology, Physics Division, University of Camerino, via Madonna delle Carceri, 9, I-62032 Camerino (MC), Italy, and INFN, Sezione di Perugia, Italy}
\date{\today}

\begin{abstract}
We describe a scheme for entangling mechanical resonators which is efficient also beyond the resolved sideband regime. It employs the radiation pressure force of the squeezed light produced by a degenerate optical parametric oscillator, which acts as a reservoir of quantum correlations (squeezed reservoir), and it is effective when the spectral bandwidth of the reservoir and the fields frequencies are appropriately selected. 
It allows for the steady state preparation of mechanical resonatrs in entangled EPR states
and can be extended to the preparation of many entangled pairs of resonators which interact with the same light field, in a situation in which the optomechanical system realizes a star-like harmonic network.
\end{abstract}

\maketitle

\section{Introduction}

Micro and Nano-Mechanical resonators represent a promising platform for the investigation of macroscopic quantum mechanical phenomena which involve the collective dynamics of a large number of quantum constituents. These studies are interesting, on the one hand, for the study of the boundary between classical and quantum realm, and on the other for possible quantum technology applications~\cite{Genes,Aspelmeyer}. Quantum behavior of mechanical degrees of freedom, such as squeezing~\cite{Wollman,Pirkkalainen} and the entanglement with light fields~\cite{Palomaki}, have been already observed. Much effort is currently devoted to the preparation of entangled mechanical systems~\cite{Mancini02,Hartmann,Pirandola06,Muller-Ebhardt08,Borkje,Wang,Woolley14,Li15,Eisert04}. 
Among the many proposals, a few have suggested the use of squeezed light as a convenient reservoir of quantum correlations which could be transferred to the mechanical elements in order to entangle them~\cite{Zhang03,Pinard,Huang,Yang}.
Squeezed light is actually used as a powerful tool to manipulate atomic, optical, mechanical and biological systems, and it has allowed, for example, for the squeezing of the collective spin of a gas of atoms~\cite{Hald}, for the modification of the radiation properties of artificial atoms~\cite{Murch} and for enhancing the sensitivity of detection devices~\cite{LIGO13, Steinlechner,Baune,Clark,Lucivero,Taylor}. Squeezed light has been also proposed as an efficient tool to entangle arrays of quantum systems by driving one or a few of its elements~\cite{ZippilliPRL13,ZippilliPRA14,ZippilliPRA15}.

In this context, here we show that the squeezed light, generated by a degenerate parametric oscillator operating below threshold, can be exploited to efficiently drive two mechanical resonators, which are coupled to a single mode of an optical cavity, into a two-mode squeezed state (EPR entanglement), also when the system is operating beyond the resolved sideband regime, i.e. when the cavity linewidth is of the order of the mechanical frequency (a regime that is typically considered inefficient for quantum manipulations of optomechanical systems). This is achieved by properly selecting the bandwidth of the squeezed field and the fields frequencies. In our analysis we employ a description which takes into accounts the finiteness of the spectral bandwidth of the squeezed reservoir, similar to Refs.~\cite{Jahne,ZippilliPRA14}, and, thereby, we demonstrate that the engineering of the spectral properties of the reservoir can be instrumental to the realization of quantum-coherent dynamics. 

In this analysis we also clarify the role of the fields frequencies for the establishment of the steady state entanglement, and we show that this protocol can be made more efficient than 
other similar protocols, also in the resolved sideband regime, by properly tuning the fields frequencies. Specifically, differently from Ref.~\cite{Huang} that describes the squeezing of a single collective mode of the two resonators (while the orthogonal one remains decoupled from the light), here we show that both collective modes can be efficiently squeezed hence realizing a highly pure state with genuine EPR correlations which exhibit significantly larger entanglement.
Correspondingly, we show that this result is achieved with a single optical cavity resulting in a protocol that is significantly simpler and easier to be realized experimentally as compared to the proposal of Ref.~\cite{Pinard}.

Our scheme can be also generalized to the  preparation of many pairs of entangled mechanical resonators which interact with a single optical cavity. These results are reminiscent of those described in~\cite{ZippilliPRA15}. However, while in~\cite{ZippilliPRA15} the steady state entanglement of many pairs is discussed for the case of a chain of harmonic oscillators, here we show that analogous results can be achieved also for a different topology of the quantum array. In fact, the optomechanical system realizes a star-like network of oscillators where the central one is the optical cavity, which is driven by the squeezed reservoir, and the mechanical resonators constitute the external nodes which get entangled in pairs. This observation could pave the way to possible future investigations of the role of the array topology on the entanglement which can be extracted from a squeezed reservoir.

The paper is organized as follows. In Sec.~\ref{intro} we introduce the model and discuss the conditions for the mechanical entanglement. In Sec.~\ref{secEPR} we describe the numerical results for the stationary EPR mechanical entanglement and we show that it is non zero also beyond the resolved sideband regime. In Sec.~\ref{section_stst} we analyze the effect of various noise sources on the steady state entanglement and its sensitivity to variations of other system parameters. The possibility to extend this scheme for the preparation of many pairs of entangled mechanical resonators which interact with a single optical cavity is discussed in Sec.~\ref{2N_memb}, and finally Sec.~\ref{con} is for conclusion.   

\begin{figure}[!tb]
\includegraphics[width=8.5cm]{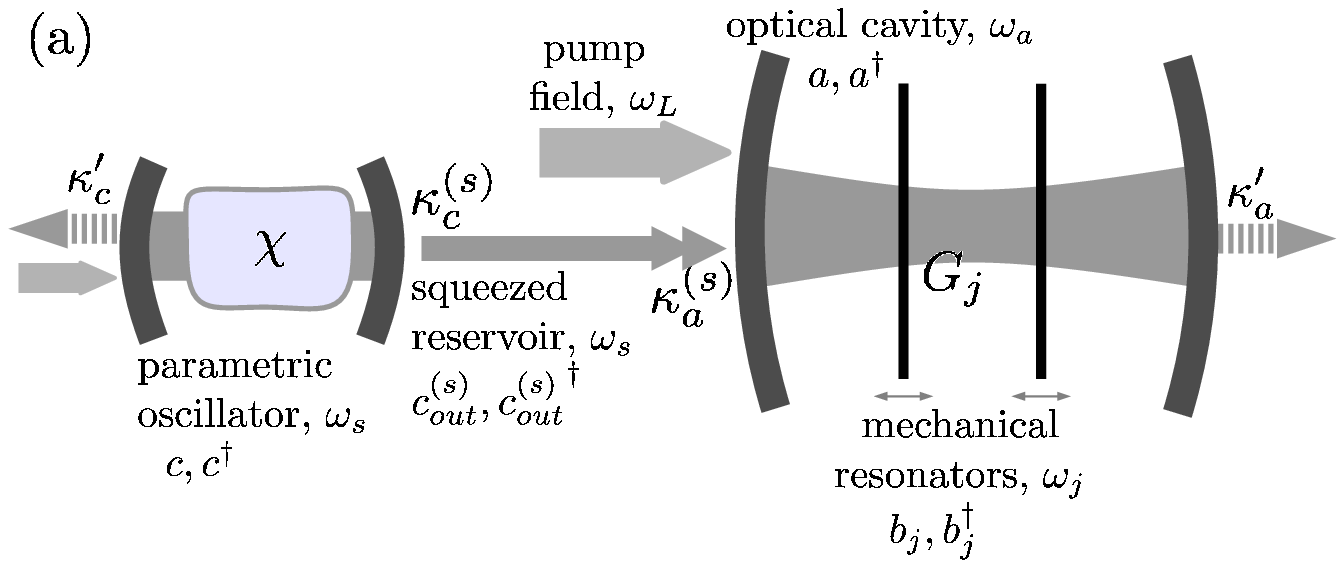}
\includegraphics[width=7.5cm]{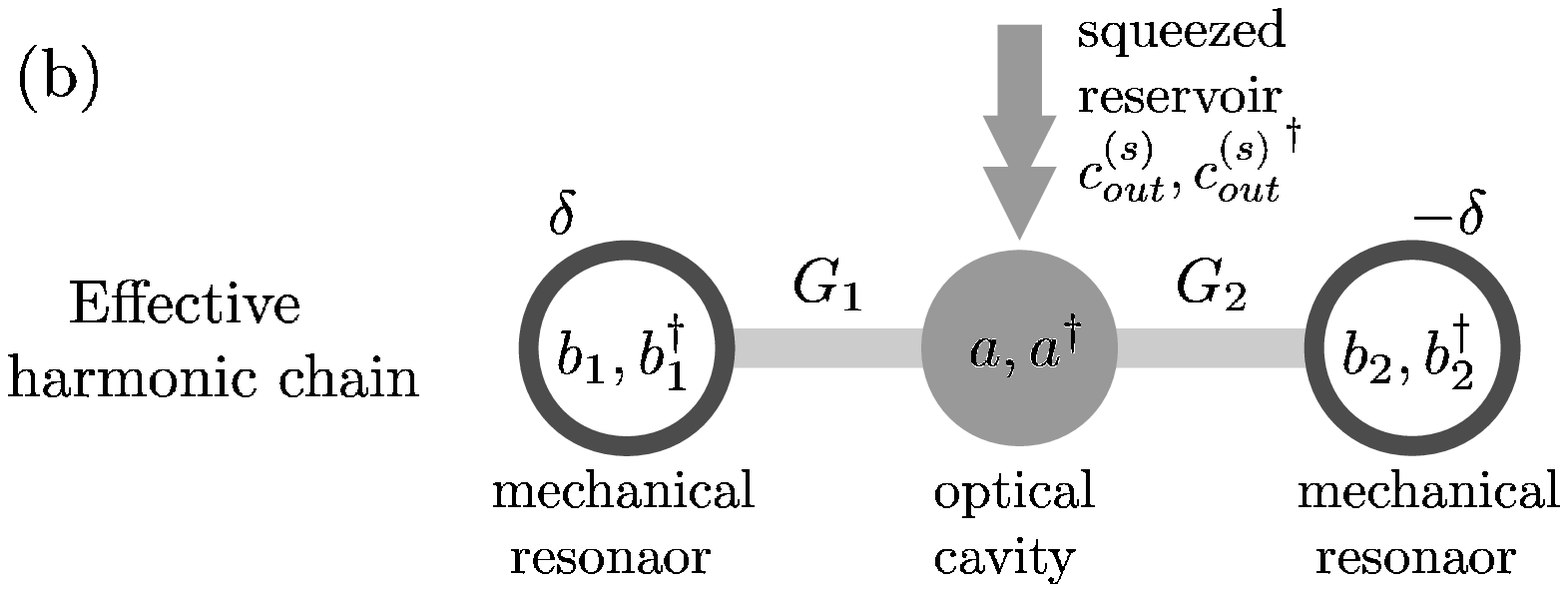}
\caption{(a) The system: The squeezed output field of a parametric oscillator drives an optomechanical system with two mechanical resonators. (b) Effective chain of harmonic oscillators realized by the optical cavity and the two mechanical resonators.}\label{fig_scheme}
\end{figure}

\section{Model and Discussion}\label{intro}

\begin{figure*}[!thb]
\includegraphics[width=18cm]{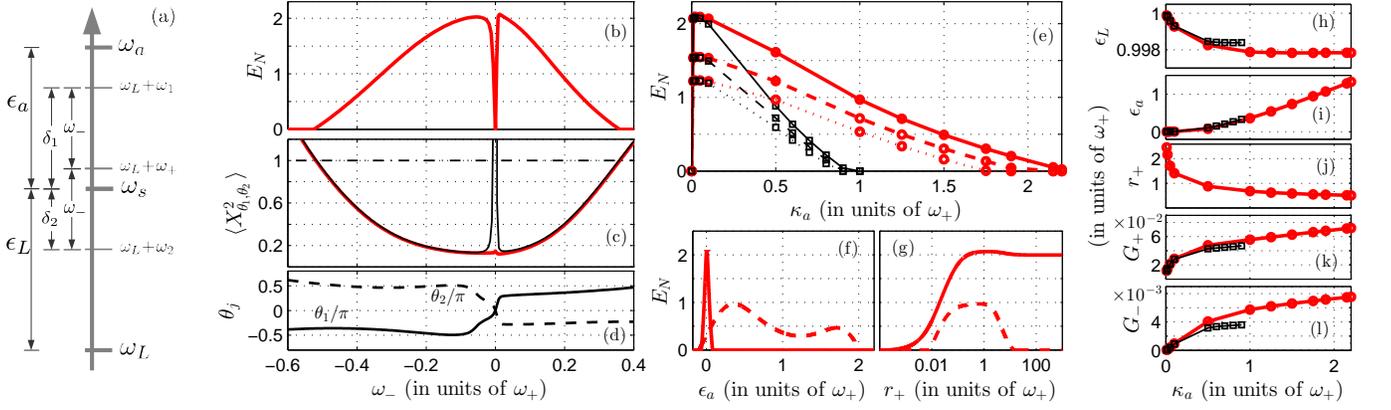}
\caption{
(a) Frequency configuration.  
(b) Logarithmic negativity $E_N$, (c) two-mode squeezing variance $\av{X_{\theta_1,\theta_2}^2}$ of the maximum squeezed collective mechanical variable (thick red curve) and of the orthogonal one $\av{X_{\theta_1+\pi/2,\theta_2-\pi/2}^2}$ (thin black curve), and (d) corresponding phases $\theta_j$ as a function of the mechanical frequency difference $\omega_-$. The driving field is squeezed by $\sim\!10$dB below the vacuum noise over a bandwidth of $r_+=1.4\omega_+$. The other parameters are $\chi=0.5\omega_+$, $\kappa_c=0.9\omega_+$, $\kappa_c'=0$, $\kappa_a=0.1\omega_+$, $\kappa_a'=0$, $G_1=0.03\omega_+$, $G_2=0.03\omega_+$, $\epsilon_L=\omega_+$, $\epsilon_a=4\times10^{-3}\omega_+$, $\gamma_1=\gamma_2=0.5\times10^5\omega_+$ and $n_{T1}=n_{T2}=10$. In (c) the dashed-dotted line indicates the vacuum noise level.
(e) $E_N$, at $\omega_-=0.01\omega_+$, as a function of the cavity linewidth $\kappa_a$,  for (solid lines) $\kappa_a'=0$ and (dashed and dotted lines) $\kappa_a'=0.1\kappa_a$, and for (solid and dashed lines) $\kappa_c'=0$ (i.e. $10$dB external squeezing) and (dotted lines) $\kappa_c'=0.1\kappa_a$ (i.e. $7.6$dB external squeezing). The red lines (circles) are evaluated with the full model in Eq.~\rp{linQLE}, and the black lines (squares) assuming an infinite bandwidth reservoir with the same value of squeezing at the central frequency. (f) and (g) $E_N$ evaluated with the full model as a function of, respectively, $\epsilon_a$ and $r_+$, for $\kappa_a'=\kappa_c'=0$, (solid lines) $\kappa_a=0.1\omega_+$, (dashed lines) $\kappa_a=\omega_+$, and for the other parameters equal to those used for the corresponding points in (e).
Each point in (e) has been optimized over $\epsilon_L$, $\epsilon_s$, $r_+$ and $G_\pm=\pt{G_1\pm G_2}/2$. The corresponding values are shown in (h)-(l). 
The other parameters in (e)-(l) are as in plot (b).
}\label{fig_res}
\end{figure*}

We first consider two mechanical resonators, at frequencies $\omega_{j}$ for $j=1,2$ and which are damped at rate $\gamma_j$. They interact by radiation pressure with a resonant mode of an optical cavity at frequency $\omega_a$, linewidth $\kappa_a$, and which is simultaneously driven by a pump laser field at frequency $\omega_L$ and by a squeezed field, with central frequency $\omega_s$ [see Fig.~\ref{fig_scheme} (a)]. The squeezed field is generated by a degenerate parametric oscillator arranged in a cascade configuration~\cite{Gardiner93,Carmichael} with the optomechanical system. In our scheme the pump frequency is always smaller then the cavity frequency and the resonators are sufficiently weakly coupled to the cavity field hence assuring the stability of the system (see App.~\ref{appStSt} for details). In this case the optomechanical dynamics can be linearized about the average cavity field and mechanical oscillations, such that the quantum Langevin equations for the fluctuations, described by the annihilation and creation operators of cavity photons, $a$ and $a\da$, and of  mechanical excitations, $b_j$ and $b_j\da$, in a reference frame in which the state of the cavity mode is rotating at the squeezed field frequency $\omega_s$ and that of the mechanical resonators at the detuning frequency between pump and squeezed fields $\epsilon_L=\omega_s-\omega_L$, are given by
\begin{eqnarray}\label{linQLE}
\dot{a}&=& -\pt{\kappa_{a}+\ii\epsilon_a}\, a+\ii\sum_{j=1}^2 G_j\pt{b_j+b_j\da\,\ee^{2\ii\,\epsilon_L\,t}}
+\sqrt{2\kappa_a}\, a_{in}
\\
\dot{b}_j&=&-\pt{\frac{\gamma_j}{2}+\ii\,\delta_j}b_j+\ii \pt{G_j^*\, a+G_j\, a^\dagger\,\ee^{2\ii\,\epsilon_L\,t}}
+\sqrt{\gamma_j}\,b_{j,in},\nn 
\end{eqnarray}
where $\epsilon_a=\omega_a-\omega_s$ is the detuning between cavity and squeezed field, $\delta_j=\omega_j-\epsilon_L$ are the mechanical detunings [see Fig.~\ref{fig_res} (a)], $G_j$ are the linearized optoemchanical coupling strengths which are proportional to the pump field, and $b_{j,in}$ are the delta correlated mechanical noise operators which account for the mechanical effects of the thermal environment at temperature $T$, such that $\pq{b_{j,in}(t),b\da_{j,in}(t')}=\delta(t-t')$ and $\langle b_{j,in}(t) b_{j',in}\da(t')\rangle = \delta_{j,j'}\,\delta(t-t')\,(n_{T_j}+1)$, with  $n_{T_j}=(\ee^{\hbar \omega_{j}/K_B T}-1)^{-1}$ the mean thermal occupation number of the mechanical modes. Finally the noise operator $a_{in}$ accounts for the effect of the external electromagnetic environment and its correlation functions are determined by the output field of the degenerate parametric oscillator, that is characterized by the non-linear self-interaction strength $\chi$ and the cavity linewidth $\kappa_c$, and whose annihilation and creation operators $c$ and $c\da$, fulfill the equation $\dot c=-\kappa_c\,c+\chi\,c\da+\sqrt{2\kappa_c}c_{in}$. Apart from the environmental modes which are squeezed by the parametric oscillator and that are controlled to drive the optomechanical system, other uncontrolled external vacuum modes of the electromagnetic field induce additional optical losses at rate $\kappa_a'$ and $\kappa_c'$ to the cavities of, respectively, the optomechanical system and the parametric oscillator. In our model, the corresponding noise can be taken into account by considering input noise operators of the form 
\begin{eqnarray}
z_{in}=\frac{\sqrt{\kappa_z\al{s}}\,z_{in}\al{s}+
\sqrt{\kappa_z'}\,z_{in}'}{\sqrt{\kappa_z}} \ \ \ {\rm for}\ \  z\in\pg{a,c}
\end{eqnarray}
where $\kappa_z\al{s}=\kappa_z-\kappa_z'$ for $ z\in\pg{a,c}$ are the rates at which photons are exchanged between the cavities and the squeezed field.
In particular, $z_{in}$ has been decomposed as the sum of two uncorrelated bosonic operators: $z_{in}\al{s}$ which is related to the external squeezed modes, and $z_{in}'$ for residual uncontrolled vacuum modes of the electromagnetic field. Note that here $\kappa_z$ is the total decay rate so that $\kappa_z'\leq\kappa_z$. The delta correlated input noise operator $a'_{in}$, $c'_{in}$ and $c\al{s}_{in}$ obey the relation $\av{a'_{in}(t)\,{a'_{in}}\da(t')}=\av{c'_{in}(t)\,{c'_{in}}\da(t')}=\av{c\al{s}_{in}(t)\,{c\al{s}_{in}}\da(t')}=\delta(t-t')$ with zero cross-correlations. The operator for the input squeezed field is instead equal to the bosonic annihilation operator for the output field of the parametric oscillator $a_{in}\al{s}=c_{out}\al{s}$, which fulfills the standard relation $c_{out}\al{s}=\sqrt{\kappa_c\al{s}}\,c-c_{in}\al{s}$. When operating below threshold, the steady state of the parametric oscillator is characterized by the correlation functions 
\begin{eqnarray}\label{corr}
\av{c_{out}\al{s}(t)\,{c_{out}\al{s}}\da(t')}&=&\delta(t-t')+v_-(t-t')
\nn\\
\av{c_{out}\al{s}(t)\,c_{out}\al{s}(t')}&=&v_+(t-t')
\end{eqnarray}
where we have introduced the functions $v_-(\tau)$ and $v_+(\tau)$ 
that determine, respectively, the number of excitations and the strength of the field self-correlations, and are given by 
\begin{eqnarray}\label{nm}
v_\pm(\tau)=\frac{\chi\,\pt{\kappa_c-\kappa_c'}}{2}
\pq{\frac{ \ee^{-r_-\abs{\tau}}}{r_-}\pm\frac{ \ee^{-r_+\abs{\tau}}}{r_+}
}
\end{eqnarray}
with $r_\pm=\kappa_c\pm\chi$. The parameter $r_+$ is actually the decay rate of the correlation function of the maximum squeezed quadrature of the field, namely it is the squeezing bandwidth, and similarly $r_-$ is the decay rate of the correlations of the anti-squeezed quadrature (see App.~\ref{appStSt}). The expression for $\nu_\pm(t)$ can be used to determine a closed set of equations for the correlation functions of the optomechanical system, from which it is possible to determine the system steady state (see App.~\ref{appStSt}). 

Steady state mechanical entanglement is obtained when 
$$\delta_1\sim-\delta_2\ ,$$ 
namely when the detuning between pump and squeezed field are close to the mechanical average frequency $\epsilon_L\sim\omega_+$, such that $\delta_1\sim-\delta_2\sim\omega_-$, with $$\omega_\pm=\frac{\omega_1\pm\omega_2}{2}$$ [see Fig.~\ref{fig_res} (a)].  
Under this condition, the squeezed reservoir induces the cooling of a specific effective chain of Bogoliubov modes of the three oscillators (the optical cavity plus the two mechanical resonators) the vacuum state of which is, indeed, characterized by a two-mode squeezed state for the mechanical resonators (see App.~\ref{appStSt_ideal}). 
Similar results have been already discussed in~\cite{ZippilliPRA15}, where it is shown that a broadband squeezed reservoir can drive a chain of linearly coupled harmonic oscillators to a steady state featuring a series of nested entangled pairs when only the central oscillator is driven by the reservoir. In the present case the optomechanical system realizes a minimal chain of three oscillators, as depicted in Fig.~\ref{fig_scheme} (b), where the central one is the optical cavity and the other two are the mechanical resonators which are driven to the entangled steady state. 

Strong entanglement can not be achieved when the mechanical resonators are degenerate $\omega_-=0$. In this case, in fact, the system has a mechanical dark mode, namely a normal mode with only mechanical components, which, hence, remains unaffected by the light (see also Ref.~\cite{Huang}). The condition of opposite detunings is analogous to the one identified in Ref.~\cite{Woolley14}, where the cooling of mechanical Bogoliubov modes is induced by two-frequency drives.  
In the present case, the steady state entanglement results from the transfer of quantum correlations from the squeezed reservoir to the mechanical resonators, hence it can be achieved efficiently when only the red sideband transitions, which describe exchange of excitations between vibrations and light, are relevant.
Non-resonant blue sideband transition, described by the time dependent terms in Eq.~\rp{linQLE}, are instead detrimental effects which degrade the transfer dynamics and, thus, should be negligible.

Typically, in order to select specific mechanical processes one works in the resolved sideband regime that is characterized by a large mechanical frequency $\omega_j\gg G_j,\kappa_a$. Under this condition, optimal entanglement is obtained when $\epsilon_a\sim0$ (that together with the condition of opposite detunings implies $\omega_a-\omega_L\sim\omega_+$). In this case the system dynamics is accurately described by a model with an infinite bandwidth reservoir achieved for $r_+\to\infty$ (corresponding to a Lindbland master equation with time independent coefficients). However, we remark that, in reality efficient entanglement is obtained whenever
$r_+\gg \delta_j,G_j,\gamma_j\,n_{Tj}$, that is when $r_+$ is much larger than the the relevant band of frequency of the optoemchanical dynamics which is characterized by only the mechanical detunings $\delta_j$, the damping rates $\gamma_j\,n_{Tj}$ and the couplings $G_j$.

Beyond the resolved sideband regime, instead, the standard description of a broadband reservoir would predict no entanglement. 
Here we demonstrate that, by properly selecting the squeezing bandwidth and the field frequencies, our scheme works efficiently also when $\kappa_a\gtrsim \omega_+ $.
While the cooling of the Bogoliubov modes always requires $\delta_1\sim-\delta_2$ (i.e $\epsilon_L\sim\omega_+$), in this case, in order to reduce the effect of blue sideband transitions, it results convenient to reduce the pump frequency, such that $\omega_a-\omega_L>\omega_+$,  as a compromise between efficient driving and reduction of blue sideband processes, in a way similar to the optimization of laser cooling with large cavity linewidth. Then, the condition of opposite mechanical detunings can be maintained by shifting correspondingly also the frequency of the squeezed field $\omega_s$ so that also $\epsilon_a$ increases. The effect of the blue sideband transitions can be further mitigated, and strong entanglement can be achieved, by reducing the reservoir photons at the blue sideband frequency, namely by using squeezed reservoir with a finite bandwidth which should be not much larger than the mechanical frequencies.

\section{Results: Stationary mechanical EPR entanglement}\label{secEPR}

The results that we have just discussed are described by Fig.~\ref{fig_res} (b)-(l).
We characterize the steady state entanglement in terms of the Logarithmic negativity $E_N$ for the two resonators. It is a measure of bipartite entanglement that, in the case of Gaussian states, can be computed by standard techniques in terms of the correlation matrix of the system (see for example the appendix in Ref.~\cite{ZippilliNJP15} for details). In (b) we plot $E_N$ as a function  of $\omega_-$. We observe a wide region of strong entanglement, of the order of the average mechanical frequency $\omega_+$, limited by blue sideband transitions which become relevant at large $\omega_-$, and that drops to zero at $\omega_-=0$ as a consequence of the presence of the mechanical dark mode [the one describing the relative motion in the case of Fig.~\ref{fig_res} (b)-(d)].
In Fig~.\ref{fig_res} (c) we plot the two-mode squeezing variance $\av{X_{\theta_1,\theta_2}^2}$ of the maximum squeezed collective mechanical variable $X_{\theta_1,\theta_2}=\sum_j\pt{b_j\ee^{\ii\theta_j}+b_j\da\ee^{-\ii\theta_j}}/\sqrt{2}$ (thick red curve) and of the orthogonal one $\av{X_{\theta_1+\pi/2,\theta_2-\pi/2}^2}$ (thin black curve) as a function of the mechanical frequency difference $\omega_-$. The corresponding phases $\theta_j$ are reported in Fig.~\ref{fig_res} (d). Specifically, plot (c) shows that at finite $\omega_-$, the observed entanglement is of EPR type with the two collective mechanical variables which are squeezed below the vacuum noise  level, and the maximum achievable squeezing is essentially equal to the squeezing of the reservoir. Close to the degenerate case $\omega_-\sim 0$, instead, only one variable can be efficiently squeezed because the other, that corresponds to the dark mode, is decoupled from the light.
Plot (e) shows how the results corresponding to the finite bandwidth (circles) exhibit always larger $E_N$ as compared to those obtained in the infinite bandwidth limit (squares). In particular significant values of the logarithmic negativity are observed also for $\kappa_a\gtrsim\omega_+$, when the latter model predicts no entanglement. Here we also show that when additional uncontrolled optical losses ($\kappa_a'\neq0$ and $\kappa_c'\neq0$) are taken into account the value of $E_N$ is reduced but the overall entanglement dynamics is still in order.
These results are computed by optimizing $E_N$, at each value of $\kappa_a$, over the squeezing bandwidth, the fields frequencies and the optomechanical couplings, the specific values of which are reported in the plots (h)-(l).
(h) shows that $\epsilon_L\sim\omega_+$ (i.e. $\delta_1\sim-\delta_2$) is always the optimal condition. (i), instead shows that $\epsilon_a$ should increase correspondingly. We also note that, as depicted in (f), the entanglement is more stable, against variation of $\epsilon_a$, when $\kappa_a$ is large (dashed line). The finite value of the optimal bandwidth is reported in (j) and correspondingly (g) shows that at small $\kappa_a$ (solid line) maximum $E_N$ is observed for sufficiently large $r_+$, while at large $\kappa_a$ (dashed line) the optimal range of $r_+$ shrinks around a value of the order of the mechanical frequency. Finally the value of the optomechanical couplings are reported in (k) and (l). Optimal $E_N$ is obtained when the couplings are essentially equal as in (l) (similar to the results in Ref.~\cite{ZippilliPRA15}), and their values should be sufficiently small to reduce the effect of the non-resonant blue sideband processes [see (k)].

\section{Sensitivity of the protocol to noise and to variations of other parameters}\label{section_stst}

\begin{figure}[b!t]
\includegraphics[width=6cm]{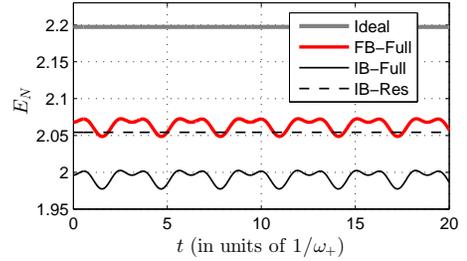}
\caption{Steady state logarithmic negativity $E_N$ as a function of time. The plot displays residual oscillation with frequency $\sim\!2\omega_+$. The driving field is squeezed by $\sim\!10$dB below the vacuum noise level over a bandwidth of $r_+=1.4\omega_+$. The inset specify how each curve has been evaluated: ``Ideal" stands for the ideal limit evaluated with Eq.~\rp{EN0}; ``FB-full" indicates the results evaluated with the full model in Eq.~\rp{stst_full_fbw} which takes into account a finite bandwidth reservoir; ``IB-full" instead corresponds to a similar model [Eq.~\rp{stst_full_ibw}] which assumes an infinite bandwidth reservoir; finally ``IB-res" corresponds to the approximate results of Eq.~\rp{stst_resonant_ibw} where non-resonant blue sideband transitions are neglected and the reservoir has infinite bandwidth. The other parameters are $\chi=0.5\omega_+$, $\kappa_c=0.9\omega_+$, $\kappa_c'=0$,  $\kappa_a=0.1\omega_+$, $\kappa_a'=0$, $G_1=0.03\omega_+$, $G_2=0.03\omega_+$, $\omega_-=0.01\omega_+$, $\epsilon_L=\omega_+$, $\epsilon_a=4\times10^{-3}\omega_+$, $\gamma_1=\gamma_2=0.5\times10^5\omega_+$ and $n_{T1}=n_{T2}=10$.
}\label{fig_t}
\end{figure}
\begin{figure*}[b!ht]
\includegraphics[width=13cm]{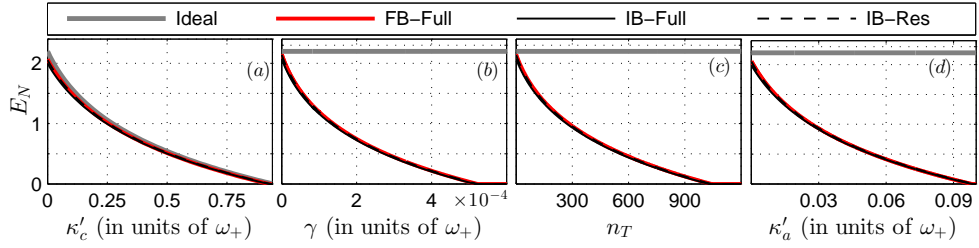}
\caption{Logarithmic negativity as a function of (a) $\kappa_c'$, (b) $\gamma=\gamma_1=\gamma_2$, (c) $n_T=n_{T1}=n_{T2}$ and (d) $\kappa_a'$. The other parameters and the line styles are as in Fig.~\ref{fig_t}.
}\label{fig_noise}
\end{figure*}
\begin{figure*}[t!b]
\includegraphics[width=16cm]{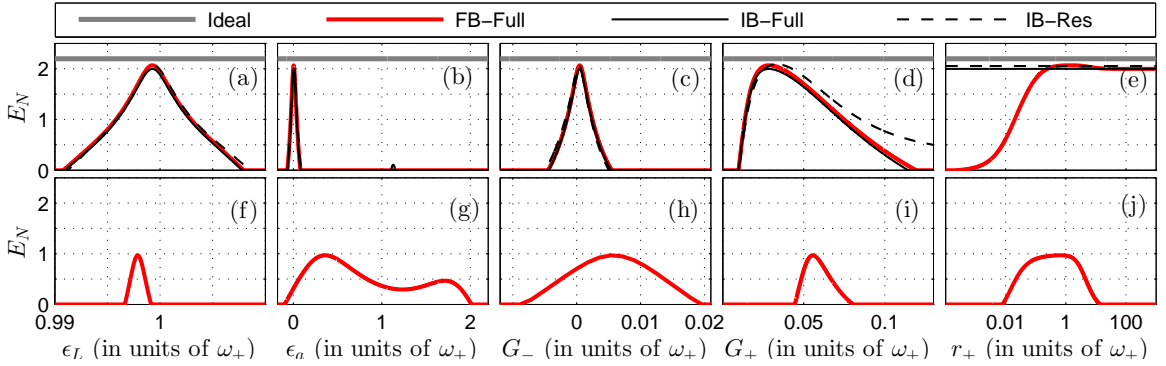}
\caption{Logarithmic negativity, for (a)-(e) $\kappa_a=0.1\omega_+$ and (f)-(j) $\kappa_a=\omega_+$, as a function of (a), (f) $\epsilon_L$, (b), (g) $\epsilon_a$, (c), (h) $G_-=\pt{G_1-G_2}/2$, (d), (i) $G_+=\pt{G_1+G_2}/2$ and (e), (j) $r_+$ with fixed ratio $r_-/r_+=1/3$ which fixes the value of the squeezing at the central frequency of the driving field.
The other parameters for the plots (a)-(e) are equal to the values corresponding to the result at $\kappa_a=0.1\omega_+$ and $\kappa_a'=0$ in Fig.~\ref{fig_res} (e), and the other parameters for plots (f)-(j) are equal to the values corresponding to the result at $\kappa_a=\omega_+$ and $\kappa_a'=0$ in Fig.~\ref{fig_res} (e). The line styles are as in Fig.~\ref{fig_t}. In the plots (f)-(j) the thin solid black lines 
are not present because the full model with infinite bandwidth reservoir predicts no entanglement; The thin dashed black lines are, instead, not present because the approximate model which neglect the non resonant processes is not valid for the value of $\kappa_a$ used in these plots.
}\label{fig_opt}
\end{figure*}

In this section we provide additional analysis of the steady state entanglement. In particular we investigate in detail its dependence on the various noise sources, and we study its sensitivity to variations in the field frequencies, the optomechanical couplings and the squeezing bandwidth.

In order to gain insight in the system dynamics and to understand the origin of the various behavior that we describe hereafter we compare, in Figs.~\rp{fig_t}, \rp{fig_noise} and \rp{fig_opt}, the results obtained with the full model [solid thick red curves evaluated with  Eq.~\rp{stst_full_fbw}], with other achieved with approximated formulas which are valid under specific conditions. Specifically, we also consider the results obtained with an infinite bandwidth squeezed reservoir [solid thin black curves evaluated with Eqs~\rp{stst_full_ibw}], and the results obtained when non-resonant blue-sideband transition processes are also neglected [dashed thin black lines, evaluated with Eq.~\rp{stst_resonant_ibw}]. Finally we also compare these results with the ideal limit (solid gray lines), evaluated with Eq.~\rp{EN0}, that corresponds to negligible optical losses and thermal noise. 

\subsection{Residual oscillations}

In Fig.~\ref{fig_t}, we observe that the steady state solution exhibits residual oscillations, that are a result of the blue sideband transitions. This can be understood by comparing the results for the full model (thick red curve)
with the dashed thin curve, obtained when neglecting non-resonant terms, that instead shows no oscillations. These oscillations are very small when $G_j\ll\omega_j$, which is the regime of interest here. For this reason and without loss of generality, the results presented in Sec.~\ref{secEPR} and those described in the following sections have been evaluated for the  specific time at which the phase of the oscillating terms is zero [namely we have set $t=0$ in Eqs.~\rp{stst_full_fbw} and \rp{stst_full_ibw}]. We also note that Fig.~\ref{fig_t} is obtained in the resolved sideband regime ($\kappa_a\ll\omega_+,G_j$), and as expected the three numerical results [evaluated with Eqs.~\rp{stst_full_fbw}, \rp{stst_full_ibw} and \rp{stst_resonant_ibw}] are very close. The small differences are due to residual effects of non-resonant blue sideband transitions, which reduce the maximum achievable entanglement when the squeezing bandwidth is large as compared to the mechanical frequencies (see the solid thin black curve obtained with an infinite bandwidth reservoir). Finally we observe that, due to the high quality factor of the resonators and the relatively low number of thermal excitations $n_{Tj}$, these curves are significantly close to the ideal limit set by the thick gray line.

\subsection{Effect of noise}  
  
The strength of the achievable steady state entanglement is ultimately limited by the amount of squeezing of the driving field, by the mechanical thermal noise (at rate $\gamma_j\, n_{Tj}$) and by residual losses of the cavity field (at rate $\kappa_a'$). 

Also the plots in Fig~\ref{fig_noise} have been evaluated in the resolved sideband regime and, as a result, the three curves evaluated with Eqs.~\rp{stst_full_fbw}, \rp{stst_full_ibw} and \rp{stst_resonant_ibw} are always very close to each other and barely distinguishable.  

Fig~\ref{fig_noise} (a) report the steady state logarithmic negativity $E_N$ as a function of $\kappa_c'$, namely the loss rate into uncontrolled output electromagnetic modes of the parametric oscillator. When $\kappa_c'$ is finite not all the squeezing produced by the parametric oscillator is efficiently transferred to the optomechanical system. Hence, as this quantity is increased, the amount of squeezing of the reservoir, that actually drives the optomechanical system, is reduced and correspondingly the achievable entanglement decreases. In any case, as far as the reservoir is actually squeezed, the entanglement dynamics is still in order and the value of $E_N$ is bounded by the amount of squeezing of the reservoir as described by Eq.~\rp{EN0} and as shown by the fact that all the curves are always very close to the ideal value described by the solid thick grey curve. Figs.~\ref{fig_noise}(b) and (c), instead, show that the system is close to the ideal limit (solid thick gray line) only for sufficiently small mechanical noise (small $\gamma_j$ and $n_{Tj}$) and it decreases as either $\gamma_j$ or $n_{Tj}$ are increased. The entanglement is also limited by uncontrolled dissipation of the cavity (at rate $\kappa_a'$) as shown in Fig.~\ref{fig_noise} (d). When $\kappa_a'=0$ the external electromagnetic environment of the optical cavity is constituted only by the modes that are squeezed by the parametric oscillator and the entanglement is maximum. When, instead, part of the external modes are uncontrolled they provide additional noise which degrades the value of $E_N$.

\subsection{Behavior of the steady state entanglement against the field frequencies, the couplings and spectral properties of the reservoir}  

Here we consider high quality factor resonators, with $Q=\omega_j/\gamma_j\sim10^5$, and with $n_{T1}=n_{T2}=10$ excitations, that is the regime corresponding to the results presented in Sec.~\ref{secEPR}. In Fig.\ref{fig_opt} we report $E_N$ as a function of the fields frequencies, the optomechanical couplings and the squeezing bandwidth, for $\kappa_a=0.1\omega_+$ [plots (a)-(e)] and for $\kappa_a=\omega_+$ [plots (f)-(j)], and with the other parameters corresponding to the optimal values identified in Fig.~\ref{fig_res} (e), at the corresponding value of $\kappa_a$ when $\kappa_a'=0$. We note that the most sensitive parameter is the detuning between the driving fields $\epsilon_L=\omega_s-\omega_L$, reported in plots (a) and (f), which should be set within a relatively narrow range of frequency around the average mechanical frequencies $\omega_+$. The value of $\epsilon_a=\omega_a-\omega_a$, in plots (b) and (g), can, instead, vary over a relatively larger range especially at large values of the cavity bandwidth [the thick red solid lines in (b) and (g) are equal to the lines in Fig.~\ref{fig_res} (f)]. Another sensitive parameter is the difference between the optomechanical coupling $G_-=(\omega_1-\omega_2)/2$ in plots (c) and (h). However also in this case the system is less sensitive to $G_-$ at larger values of $\kappa_a$. On the contrary the optimal region of values of $G_+=(\omega_1+\omega_2)/2$ shrink sensibly as $\kappa_a$ increases [see plots (d) and i]. We note that although it seems reasonable that the values of $\epsilon_L$ and $\epsilon_a$ can be easily set with sufficient precision by proper tuning of the field frequencies, it may results technically challenging to set the optomechanical couplings to the desired values. Finally the system is significantly stable over the squeezing bandwidth as reported in (e) and (j). In fact large values of entanglement are observed over a wide range of values which shrink mildly, at large $\kappa_a$, around a value of $r_+$ of the order of the mechanical frequencies [the thick red solid lines in (e) and (j) are equal to the lines in Fig.~\ref{fig_res} (g)].

\begin{figure}[!tb]
\includegraphics[width=8cm]{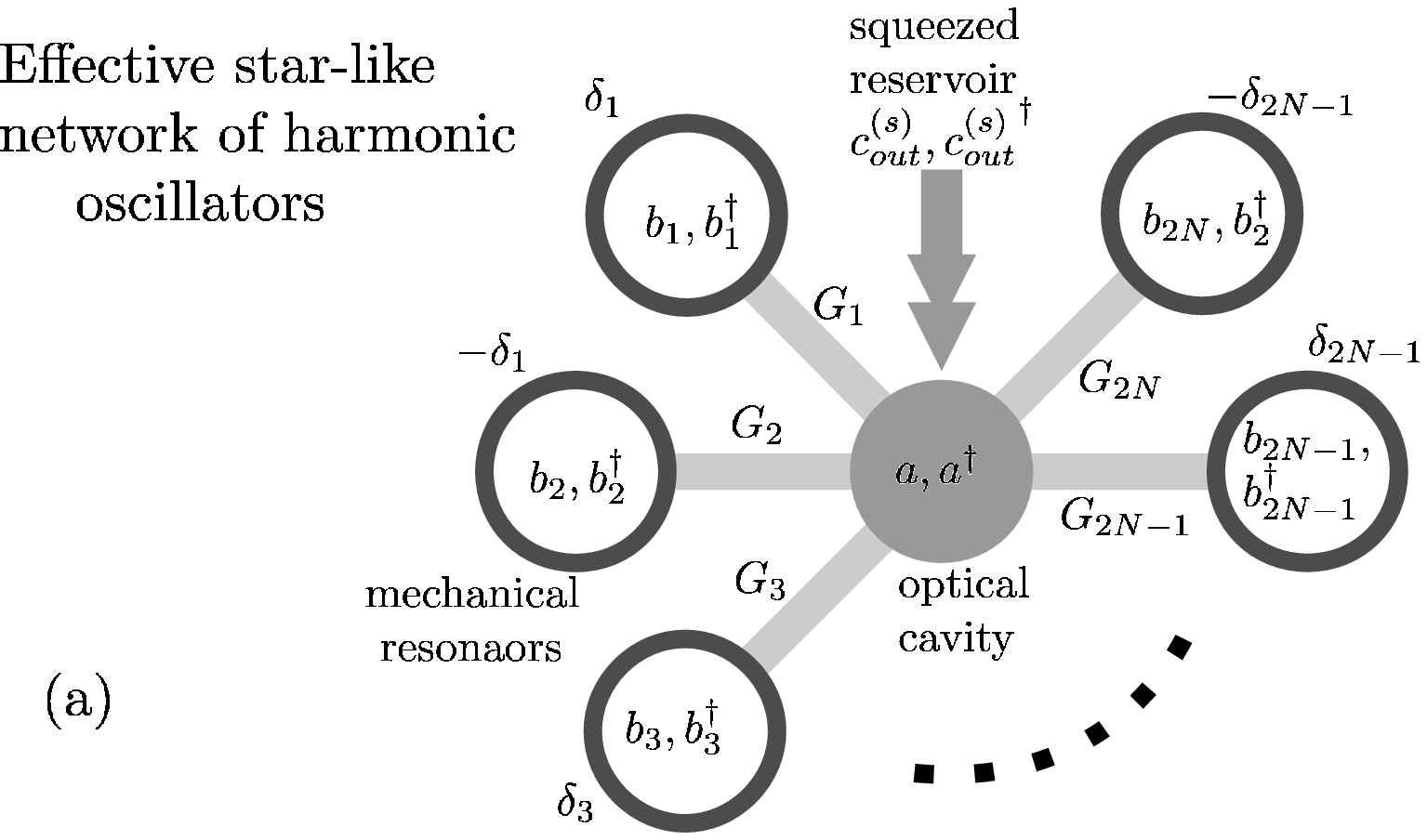}
\includegraphics[width=8.5cm]{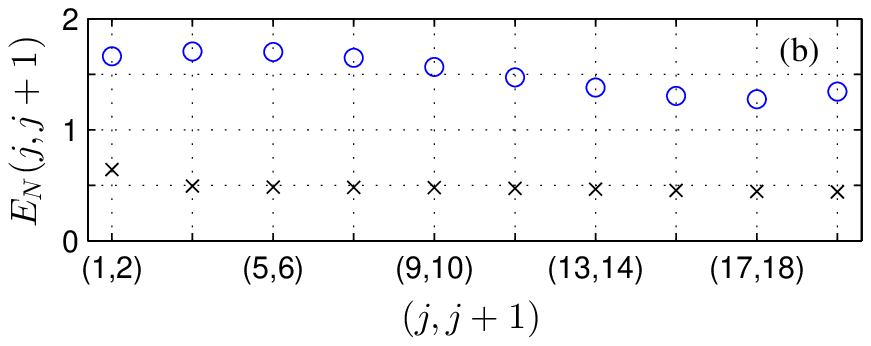}
\caption{(a) Effective star-like network realized by an optomechanical system with $2N$ mechanical resonators. (b) Logarithmic negativity $E_N(j,j+1)$ for the pairs of resonators with indices $j$ and $j+1$ in an optomechanical system with 20 mechanical resonators, i.e  $N=10$, for (circles) $\kappa_a=0.1\omega_+$, (crosses) $\kappa_a=\omega_+$ and in both cases $\kappa_a'=0$. The mechanical detunings are chosen such that $\delta_{2j-1}=-\delta_{2j}=\pq{1+3(j-1)}\delta$ with $\delta=0.01\omega_+$.  
The couplings are set to (circles) $G_{j}=0.03\omega_+, \forall j$, and (crosses) $G_{2j-1}=0.06\omega_+$ and $G_{2j}=0.05\omega_+$ for $j\in\pg{1,2\cdots N}$.
The other parameters are as in Fig.~\ref{fig_res} (b).
All the other pairs are not entangled.
}\label{fig_2N}. 
\end{figure}

\section{$\bs 2\bs\times \bs N$ mechanical resonators}\label{2N_memb}

This dynamics can be generalized to an arbitrary number of pairs $N$ of mechanical resonators inside the same optical cavity. In this case the system is described by a set of equations equal to Eq.~\rp{linQLE} but where now the sum over $j$ runs from 1 to $2N$. Also in this case each pair can be entangled if the corresponding mechanical detunings are opposite $\delta_j=-\delta_{j+1}$, and if $G_j\sim G_{j+1}$ for $j\in\pg{1,3,5\cdots}$. However the actual strength of $E_N$, will depend also on the values of the frequency difference of each pair 
$\omega_-\al{j}=\pt{\omega_j-\omega_{j+1}}/{2}$ and of the relative coupling parameters of all the pairs. In fact, similar to the case $\omega_-=0$ for two resonators, there can be situations in which the achievable entanglement is degraded due to the presence of mechanical dark modes. In particular, all the resonators can be efficiently driven into the entangled state only if none of the normal modes of the system is orthogonal to the optical cavity. Similar considerations have been already discussed in Ref.~\cite{ZippilliPRA15} for the preparation of many entangled pairs in a chain of harmonic oscillators. Here we show that analogous results are obtained for a star-like network of oscillators where the central node is the optical cavity and the external ones the mechanical resonators. We report in Fig.~\ref{fig_2N} a specific realization, with 20 mechanical oscillators, demonstrating that all and only the pairs with opposite mechanical detunings are efficiently entangled for specific values of $\omega_-\al{j}$ also beyond the resolved sideband regime.

\section{conclusion}\label{con}

In conclusion we have proved, using a realistic model which takes into account also the spectral properties of the reservoir, that entangling mechanical degrees of freedom with squeezed light is actually more efficient then what previously believed, and this effect is observable in the regime of finite bandwidth which is naturally accessible in experiments.
Specifically we have shown that the squeezed field produced by a degenerate parametric oscillator can be efficiently employed to drive mechanical resonators into EPR two-mode squeezed states. While the maximum entanglement is achieved in the resolved sideband regime, this scheme is efficient also beyond this regime if  the squeezing bandwidth and the fields frequencies are appropriately engineered. These findings and the identification of the real potentiality of squeezed light can be very relevant to the many recent experiments which explore the efficiency of using squeezed fields for the manipulation of quantum system~\cite{Murch,LIGO13, Steinlechner,Baune,Clark,Lucivero,Taylor}. We remark that these results may be beneficial also beyond the field of quantum optomechanics (superconducting nano-devices, trapped ions...) where similar dynamics could be observed.

In our case, two mechanical resonators at frequencies $2\pi\times0.99$GHz and $2\pi\times1.01$GHz, and quality factor $Q\sim 2\times 10^5$, at a temperature of $0.5$K, can be efficiently driven to an EPR entangled state featuring $\sim7$dB ($\sim3$dB) of two-mode squeezing below the vacuum noise level when coupled to an optical cavity, with linewidth of $2\pi\times0.1$MHz ($2\pi\times1$GHz) and 10\% of uncontrolled optical losses, which is fed with a field squeezed by $10$dB over a bandwidth of the order of the mechanical frequency. In a realistic scenario, in order to detect the generated entanglement one can employ additional probe fields along the lines discussed in Ref.~\cite{Li15}. Moreover this scheme is straightforwardly generalizable to situations involving many resonators, which could be relevant for the development of routers for phonon-based quantum communications and information~\cite{Habraken, Hatanaka,Fang}.
With this result we have, therefore, demonstrated that the engineering of the spectral properties of a squeezed reservoir may be useful for quantum technology applications.

\section*{Acknowledgments}

This work is supported by the European Commission through the Marie-Curie ITN cQOM and FET-Open Project iQUOEMS.

\appendix

\section{Evaluation of the steady state}\label{appStSt}

Differently form the description in the Sec.~\ref{intro}, here we study the system in a reference frame rotating at the laser field frequency $\omega_L$. In this case the linearized optomechanical dynamics is described by the equations
\begin{eqnarray}
\dot{\tilde a}&=& -\pt{\kappa_{a}+ \ii\Delta }\, \tilde a+\ii\sum_{j=1}^2 G_j\pt{\tilde b_j+\tilde b_j\da}
+\sqrt{2\kappa_a}\, \tilde a_{in}
\\
\dot{\tilde b}_j&=&-\pt{\ii\omega_{j}+\frac{\gamma_j}{2}}\tilde b_j+\ii \pt{G_j\, \tilde a^\dagger+G_j^*\, \tilde a}+\sqrt{\gamma_j}\,b_{j,in}, \nn
\label{linQLE0}
\end{eqnarray}
where $\Delta =\omega_a-\omega_L$ is the detuning between cavity and pump fields, and the operators used here are related to those in Eq.~\rp{linQLE} by the relation
\begin{eqnarray}\label{tildeab}
\tilde a&=&a\,\ee^{\ii\epsilon_L\,t}\nn\\
\tilde b_j&=&b_j\,\ee^{\ii\epsilon_L\,t}\ .
\end{eqnarray}
All the other parameters are defined as in Sec.~\ref{intro}. Moreover, while the mechanical noise operators have the same form in the two representations, in this case, an additional time dependent phase appears in the definition of the new input noise operator for the cavity field, namely
\begin{eqnarray}\label{ain}
\tilde a_{in}=\pt{\sqrt{\kappa_a\al{s}}\,c_{out}\al{s}\,\ee^{-\ii\epsilon_L\,t}+
\sqrt{\kappa_a'}\,a_{in}'}/\sqrt{\kappa_a}\ ,
\end{eqnarray}
where $c_{out}\al{s}$ is the output field of the optical parametric oscillator (OPO)
whose equations are given in the main text. We assume that the OPO operates in its steady state. Hence we can use the corresponding steady state correlation functions, defined in Eqs.~\rp{corr} and \rp{nm}, to determine a set of closed equations for the optomechanical system.
The steady state of the OPO is further characterized by the squeezing spectrum $S(\omega)$, namely the noise spectral density associated to the maximum squeezed quadrature~\cite{ZippilliNJP15}, that in the present case is $Y=c_{out}\al{s}\,\ee^{\ii\pi/2}+{c_{out}\al{s}}\da\,\ee^{-\ii\pi/2}$, 
and it is given by~\cite{WallsMilburn}
\begin{eqnarray}\label{S}
S(\omega)=\av{Y(\omega)^2}=1-\frac{4\chi\,\pt{\kappa_c-\kappa_c'}}{r_+^2+\omega^2},
\end{eqnarray}
where $\omega$ is the frequency of the spectral component of the field relative to the central frequency $\omega_s$. It shows that $r_+$ is the squeezing bandwidth of the output field~\cite{ZippilliPRA14}. Similarly one can show that the orthogonal quadrature is anti-squeezed over a bandwidth $r_-$. The limit of a broadband reservoir corresponds to a flat spectrum $S(\omega)=$const, that is obtained when $r_\pm\to\infty$.

The linearized dynamics of the fluctuations is Gaussian hence it is fully determined by their second statistical moments. In details we analyze the equation for the correlation matrix $\tilde\VV(t)=\av{\tilde \va(t)\,\tilde \va(t)^T}$, for the vector of operators $\tilde \va(t)=\pq{\tilde a(t),\tilde a\da(t),\tilde b_1(t),\tilde b\da_1(t), \tilde b_2(t),\tilde b\da_2(t)}^T$, whose equation takes the form
\begin{eqnarray}\label{VV}
\dot{\tilde{\VV}}(t)=\tilde\AAA\,\tilde\VV(t)+\tilde\VV(t)\,\tilde\AAA^T+\tilde\BB(t)
\end{eqnarray}
where $\tilde\AAA$ is the matrix of the coefficients of the system of equation~\rp{linQLE0}, given by
\begin{widetext}
\begin{eqnarray}\label{AAA}
\tilde\AAA=\left(\small\begin{array}{cccccc}
-\kappa_{a}-\ii\Delta  &0			&		\ii G_1	&	\ii G_1	&	\ii G_2	&	\ii G_2\\
0       &-\kappa_{a}+\ii\Delta  	&		 -\ii G_1^*	&	-\ii G_1^*	&	-\ii G_2^*	&	-\ii G_2\\		
\ii G_1^*	&	\ii G_1		&-\frac{\gamma_1}{2}-\ii\omega_1   & 0    & 0     &0 \\
-\ii G_1^* 	&	-\ii G_1	& 0  & -\frac{\gamma_1}{2}+\ii\omega_1  & 0 & 0  \\
\ii G_2^*	&	\ii G_2		&	0  & 0 &-\frac{\gamma_2}{2}-\ii\omega_2  & 0\\
-\ii G_2^*	&	-\ii G_2	&	0  & 0 & 0 & -\frac{\gamma_2}{2}+\ii\omega_2
\end{array}\right),
\nn
\end{eqnarray}
\end{widetext}
and the diffusion matrix $\tilde\BB(t)$ is time dependent and it is constructed in terms of the correlation matrix $\CC_{in}(t,t')=\av{\tilde \va\al{in}(t)\,{\tilde \va\al{in}(t')}^T}$ for the vector of input noise operators $\tilde \va\al{in}(t)=$ $[\,\sqrt{2\kappa_{a}}\,\tilde a_{in}(t),\sqrt{2\kappa_{a}}\,\tilde a\da_{in}(t),$ $\sqrt{\gamma_1}\,b_{1,in}(t),$ $\sqrt{\gamma_1}\,b\da_{1,in}(t),$ $  
 \sqrt{\gamma_2}\,$ $b_{2,in}(t),\sqrt{\gamma_2}\,b_{2,in}\da(t)\,]^T$, as
\begin{eqnarray}\label{BB}
\tilde\BB(t)=
\int_0^t\dd\tau\pq{
\ee^{\tilde \AAA(t-\tau)}\,
\CC_{in}(\tau,t)
+\CC_{in}(t,\tau)\,
\ee^{\tilde \AAA^T(t-\tau)}
}\ .
\end{eqnarray}
The correlation matrix $\CC_{in}(t,t')$ is determined using Eqs.~\rp{ain} and \rp{corr}, and can be decomposed into the sum of matrices with different time dependence, as
\begin{eqnarray}
&&\CC_{in}(t,t')=\delta(t-t')\,\CC\al{0}
\\&&\hspace{1cm}
+v_-(t-t')\pq{\ee^{-\ii{\epsilon_L}(t-t')}\CC\al{1,2}+\ee^{\ii{\epsilon_L}(t-t')}\CC\al{2,1}}
\nn\\&&\hspace{1cm}
+v_+(t-t')\pq{\ee^{-\ii{\epsilon_L}(t+t')}\CC\al{1,1}+\ee^{\ii{\epsilon_L}(t+t')}\CC\al{2,2}}  \nn
\end{eqnarray}
where $v_\pm$ are defined in Eq.~\rp{nm}, the matrix corresponding to the delta-correlated part of the noise is
\begin{eqnarray}
\hspace{-0.5cm}
\CC\al{0}=\pt{\small\mat{cccccc}{
0&2\,\kappa_a & 0&0&0&0\\
0&0&0&0&0&0\\
0&0&0&\gamma_1\pt{n_{T_1}+1}&0&0\\
0&0&\gamma_1\,n_{T_1}&0&0&0\\
0&0&0&0&0&\gamma_2\pt{n_{T_2}+1}\\
0&0&0&0&\gamma_2\,n_{T_2}&0
}}\nn\\
\end{eqnarray}
and the matrices $\CC\al{\ell,\ell'}$, with $\ell,\ell'\in\pg{1,2}$, are $6\times 6$ matrices with a single non-zero element, that is given by 
\begin{eqnarray}
\pg{\CC\al{\ell,\ell'}}_{\ell,\ell'}=2\, \kappa_a\al{s}\ .
\end{eqnarray}
Thereby
\begin{eqnarray}
\tilde\BB(t)&=&\CC\al{0}
+\frac{1}{2}\pq{\NN_+(t)\,\CC\al{1,2}+\CC\al{1,2}\,\NN_-(t)^T
}\nn\\&&+\frac{1}{2}\pq{
\NN_-(t)\,\CC\al{2,1}+\CC\al{2,1}\,\NN_+(t)^T}
\nn\\&&
+\frac{1}{2}\pq{\MM_+(t)\,\CC\al{1,1}+\CC\al{1,1}\,\MM_+(t)^T}\ee^{-2\ii{\epsilon_L}\,t}
\nn\\&&
+\frac{1}{2}\pq{\MM_-(t)\,\CC\al{2,2}+\CC\al{2,2}\,\MM_-(t)^T}\ee^{2\ii{\epsilon_L}\,t}
\end{eqnarray}
where
\begin{eqnarray}
\NN_\pm(t)&=&\chi\,\kappa_c\al{s}\,
\pq{
\frac{1- \ee^{-\pt{r_--\tilde\AAA\pm\ii{\epsilon_L}}t}}{r_-\pt{r_--\tilde\AAA\pm\ii{\epsilon_L}}}
-\frac{1- \ee^{-\pt{r_+-\tilde\AAA\pm\ii{\epsilon_L}}t}}{r_+\pt{r_+-\tilde\AAA\pm\ii{\epsilon_L}}}
}\nn\\
\MM_\pm(t)&=&\chi\,\kappa_c\al{s}\,
\pq{
\frac{1- \ee^{-\pt{r_--\tilde\AAA\pm\ii{\epsilon_L}}t}}{r_-\pt{r_--\tilde\AAA\pm\ii{\epsilon_L}}}
+\frac{1- \ee^{-\pt{r_+-\tilde\AAA\pm\ii{\epsilon_L}}t}}{r_+\pt{r_+-\tilde\AAA\pm\ii{\epsilon_L}}}
}\ .\nn\\
\end{eqnarray}

We are interested in the long-time solution $\tilde\VV_{st}(t)=\lim_{t\to\infty} \tilde\VV(t)$. The system is stable, hence it approaches a stationary state at large time, if all the eigenvalues of the matrix $\tilde\AAA$ have negative real part. Using the Routh-Hurwitz criterion applied to the corresponding matrix for the evolution of the quadrature operators (which is a real matrix), it is possible to show that when $\Delta >0$ this condition is equivalent to~\cite{Genes} 
\begin{eqnarray}
\Delta ^2+\kappa_a^2-2\,\Delta \sum_j\frac{G_j^2}{\omega_j}>0\ .
\end{eqnarray}
All the results discussed in this paper are obtained for parameters which fulfill this relation. By defining the linear Lyapunov operator $\tilde\LL$ which operates on a generic correlation matrix $\CC$ as
$\tilde\LL\,\CC=\tilde\AAA\,\CC+\CC\,\tilde\AAA^T$ one can formally write the solution of Eq.~\rp{VV} as 
$\tilde\VV(t)=\ee^{\tilde\LL\,t}\,\tilde\VV(0)+\int_0^t\dd\tau\,\ee^{\tilde\LL\,(t-\tau)}\,\tilde\BB(\tau)$, thus the steady state solution takes the form
\begin{eqnarray}\label{stst_full_fbw}
\tilde\VV_{st}(t)&=&-\tilde\LL^{-1}\lpg{\CC\al{0}
+\frac{1}{2}\pq{\bar\NN_+\,\CC\al{1,2}+\CC\al{1,2}\,\bar\NN_-^T
}}\\&&\rpg{+\frac{1}{2}\pq{
\bar\NN_-\,\CC\al{2,1}+\CC\al{2,1}\,\bar\NN_+^T
}}\nn\\&&
-\frac{1}{2}\pt{\tilde\LL+2\ii\,{\epsilon_L }}^{-1}\pq{\bar\MM_+\,\CC\al{1,1}+\CC\al{1,1}\,\bar\MM_+^T}\ee^{-2\ii\epsilon_L\,t}
\nn\\&&
-\frac{1}{2}\pt{\tilde\LL-2\ii\,{\epsilon_L }}^{-1}\pq{\bar\MM_-\,\CC\al{2,2}+\CC\al{2,2}\,\bar\MM_-^T}\ee^{2\ii{\epsilon_L}\,t}
\nn
\end{eqnarray}
where
\begin{eqnarray}
\bar\NN_\pm&=&\chi\,\kappa_c\al{s}\,
\pq{
\frac{1}{r_-\pt{r_--\tilde\AAA\pm\ii{\epsilon_L}}}
-\frac{1}{r_+\pt{r_+-\tilde\AAA\pm\ii{\epsilon_L}}}
}\nn\\
\bar\MM_\pm&=&\chi\,\kappa_c\al{s}\,
\pq{
\frac{1}{r_-\pt{r_--\tilde\AAA\pm\ii{\epsilon_L}}}
+\frac{1}{r_+\pt{r_+-\tilde\AAA\pm\ii{\epsilon_L}}}
}\ .\nn\\
\end{eqnarray}
Eq.~\rp{stst_full_fbw} has been used to compute numerically the thick red lines in Figs.~\ref{fig_res} (b), (e)-(l), \ref{fig_t}, \ref{fig_noise} and \ref{fig_opt}, and the lines in Fig.~\ref{fig_res} (c), (d).

\subsection{Broadband squeezed reservoir}\label{appStSt_ib}

The limit of a broadband reservoir is achieved when the decay time of the reservoir correlation functions is very short, namely when $r_\pm$ are very large, specifically much larger than the eigenvalues of the matrix $\tilde\AAA$, that is for example when $r_-\gg \kappa_a,G_j,\gamma_j$. In this limit it is legitimate to approximate the correlation functions which appears in the expression for $\BB(t)$ in Eq.~\rp{BB} as delta functions according to the relations 
\begin{eqnarray}\label{nm0t}
v_-(t)&\simeq&\delta(t)\,\bar n\ ,
\nn\\
v_+(t)&\simeq&\delta(t)\,\bar m
\end{eqnarray}
where 
\begin{eqnarray}\label{nm0}
\bar n&=& \chi\,\kappa_c\al{s}\pt{\frac{1}{r_-^2}-\frac{1}{r_+^2}} 
\nn\\
\bar m&=&\chi\,\kappa_c\al{s}\pt{\frac{1}{r_-^2}+\frac{1}{r_+^2}}
\end{eqnarray}
and $\BB(t)$ reduces to the matrix
\begin{eqnarray}
&&\tilde\BB(t)\simeq\CC\al{0}+\bar n\pt{\CC\al{1,2}+\CC\al{2,1}}
\nn\\&&\hspace{1cm}
+\bar m\,\pt{\CC\al{1,1}\,\ee^{-2\ii{\epsilon_L}\,t}+\CC\al{2,2}\,\ee^{+2\ii{\epsilon_L}\,t}}\ .
\end{eqnarray}
In this case the steady state correlation matrix can be expressed as
\begin{eqnarray}\label{stst_full_ibw}
&&\tilde\VV_{st}(t)\simeq-\tilde\LL^{-1}\pq{\CC\al{0}
+\bar n\pt{\CC\al{1,2}+\CC\al{2,1}}
}\\&&
\ \ -\bar m\pq{\pt{\tilde\LL+2\ii\,{\epsilon_L }}^{-1}\,\CC\al{1,1}\,\ee^{-2\ii{\epsilon_L }\,t}
+\pt{\tilde\LL-2\ii\,{\epsilon_L}}^{-1}\,\CC\al{2,2}\,\ee^{2\ii{\epsilon_L}\,t}}\ .
\nn
\end{eqnarray}
Eq.~\rp{stst_full_ibw} has been used to compute numerically the thin black lines in Figs.~\ref{fig_res} (e)-(l), \ref{fig_t}, \ref{fig_noise} and \ref{fig_opt}.

\subsection{Resonant processes}\label{appStSt_res}

If we also assume large mechanical frequencies such that $\Delta \gg G_j,\kappa_a,\gamma_j$ and $\epsilon_L\sim\Delta $, then it is possible to approximate the quantum Langevin equations in Eq.~\rp{linQLE0} by neglecting non-resonant terms and to obtain an equation for the correlation matrix with time independent coefficients. In order to do this, it is useful to describe the system in a reference frame rotating at the detuning frequency $\epsilon_L$, as in Sec.~\ref{intro}, where the operators in the two representations are related by Eq.~\rp{tildeab}.
Thereby the corresponding quantum Langevin equations, obtained retaining only the resonant terms (in particular we neglect blue sideband transitions), read [see Eq.~\rp{linQLE}]
\begin{eqnarray}\label{resQLE}
\dot{a}&=& -\pt{\kappa_{a}+\ii\epsilon_a}\, a+\ii\sum_{j=1}^2 G_j\, b_j
+\sqrt{2\kappa_a}\, a_{in}
\\
\dot{b}_j&=&-\pt{\frac{\gamma_j}{2}+\ii\,\delta_j}b_j+\ii \,G_j^*\, a
+\sqrt{\gamma_j}\,b_{j,in},\nn 
\end{eqnarray}
In this case the equation for the correlation matrix, $
\dot\VV(t)=\AAA\,\VV(t)+\VV(t)\,\AAA^T+\BB$,
is given in terms of the time-independent matrices
\begin{widetext}
\begin{eqnarray}\label{AAAres}
\AAA=\left(\small\begin{array}{cccccc}
-\kappa_{a}-\ii\epsilon_L & 0 & \ii G_1 & 0 & \ii G_2 & 0\\
0 &-\kappa_{a}+\ii\epsilon_L  &	0 & -\ii G_1^*	 & 0 &	-\ii G_2\\		
\ii G_1^*	& 0 &-\frac{\gamma_1}{2}-\ii\delta_1   & 0    & 0     &0 \\
0	&	-\ii G_1	& 0  & -\frac{\gamma_1}{2}+\ii\delta_1 & 0 & 0  \\
\ii G_2^*	& 0 &	0  & 0 &-\frac{\gamma_2}{2}-\ii\delta_2 & 0\\
0 &	-\ii G_2	&	0  & 0 & 0 & -\frac{\gamma_2}{2}+\ii\delta_2
\end{array}\right)\ ,
\nn
\end{eqnarray}
and
\begin{eqnarray}
&&\BB =
\pt{\small\mat{cccccc}{
2\kappa_a\al{s}\,\bar m &2\pt{\kappa_{a}+\kappa_a\al{s}\,\bar n} & 0&0&0&0\\
2\kappa_a\al{s}\,\bar n&2\kappa_a\al{s}\,\bar m&0&0&0&0\\
0&0&0&\gamma_1\pt{n_{T_1}+1}&0&0\\
0&0&\gamma_1\, n_{T_1}&0&0&0\\
0&0&0&0&0&\gamma_2\pt{n_{T_2}+1}\\
0&0&0&0&\gamma_2\,n_{T_2}&0
}}\ .\nn
\end{eqnarray}
\end{widetext}
Correspondingly the steady state solution for the correlation matrix can be formally expressed as 
\begin{eqnarray}\label{stst_resonant_ibw}
\VV_{st}=-\LL^{-1}\,\BB\ ,
\end{eqnarray}
where $\LL$ is defined by the relation $\LL\,\CC=\AAA\,\CC+\CC\,\AAA^T$. 
Eq.~\rp{stst_resonant_ibw} has been used to compute the thin dashed lines in the Figs.~\ref{fig_t}, \ref{fig_noise} and \ref{fig_opt}.

We remark that although Eqs.~\rp{stst_full_fbw} and \rp{stst_full_ibw} are evaluated in a reference frame different form the one used here [where the relation between the two is defined in Eq.~\rp{tildeab}], they can be safely used to compare the entanglement achieved in the two representation because the entanglement is invariant under local unitary operation as those defined in Eq.~\rp{tildeab}.

When the squeezed reservoir is resonant with the optomechanical cavity, i.e. $\epsilon_a=0 $, and the mechanical detunings are opposite $\delta_j=-\delta_2=\omega_-$, or equivalently $\epsilon_+=\omega_+$, then Eq.~\rp{resQLE} reproduce the model studied in~\cite{ZippilliPRA15} with only three oscillators [see Fig.~\ref{fig_scheme} (b)].

\subsection{Ideal limit}\label{appStSt_ideal}

Here we study the ideal limit in which mechanical noise and uncontrolled optical losses (respectively described by $\gamma_j\, n_{T_j}$ and $\kappa_a'$) are negligible. If we assume that the squeezed reservoir is resonant with the optomechanical cavity (i.e. $\epsilon_a=0 $), that the mechanical detunings are opposite $\delta_j=-\delta_2=\omega_-$ (or equivalently $\epsilon_+=\omega_+$) and that $G_1=G_2\equiv G$, then the system dynamics is easily described by introducing the Bogoliubov modes
\begin{eqnarray}
\hat\alpha&=&\cosh(s)\,a-\sinh(s)\,a\da
\nn\\
\hat\beta_1&=&\cosh(s)\, b_1+\sinh(s)\, b_2\da
\nn\\
\hat\beta_2&=&\cosh(s)\, b_2+\sinh(s)\, b_1\da
\end{eqnarray} 
with $\tanh(s)=(\bar n-\bar n_s)/\bar m$ and $\bar n_s=\pq{\sqrt{(2\bar n+1)^2-4\bar m}\,-1}/2$ whose quantum Langevin equations are
\begin{eqnarray}
\dot{\hat\alpha}&=& -\kappa_{a}\,\hat\alpha+\ii\, \sum_j G \,\hat\beta_j 
+\sqrt{2\kappa_a}\, \hat\alpha_{in}
\nn\\
\dot{\hat\beta}_1&=&-\ii\,\omega_-\,\hat\beta_1+\ii\, G\,\hat\alpha\ ,\nn\\
\dot{\hat\beta}_2&=&\ii\,\omega_-\,\hat\beta_2+\ii\, G\,\hat\alpha
\end{eqnarray} 
where the noise operator $\hat\alpha_{in}=\cosh(s)\, a_{in}-\sinh(s)\, a_{in}\da(t)$ corresponds to an effective thermal bath whose correlation functions are $\av{\hat\alpha_{in}(t)\,\hat\alpha_{in}\da(t')}=(\bar n_s+1)\delta(t-t')$ and $\av{\hat\alpha_{in}(t)\,\hat\alpha_{in}(t')}=0$. Thereby in this representation (and if the normal modes of the system are not orthogonal to the central oscillator~\cite{ZippilliPRA15}) the three Bogolibuv oscillators thermalize with the effective bath and approach a thermal state with $\bar n_s$ thermal excitations. Such that $\av{\hat\alpha\da\,\hat\alpha}_{st}=\bar n_s$, $=\av{\hat\beta_j\da\,\hat\beta_k}_{st}=\delta_{j,k}\bar n_s$ and $\av{\hat\alpha\,\hat\alpha}_{st}=\av{\hat\beta_j\,\hat\beta_k}_{st}=\av{\hat\alpha\da\,\hat\beta_j}_{st}=\av{\hat\alpha\,\hat\beta_j}_{st}=0$, where the index $_{st}$ indicates that the averages are performed over the steady state. Correspondingly we find that the steady state for the oscillators in the original representation is characterized by
\begin{eqnarray}
\av{ a\da\,a}_{st}&=&\bar n\ , \hspace{1.5cm}  \av{b_j\da\, b_k}_{st}=\delta_{j,k}\bar n\ ,
\nn\\
\av{ a\, a}_{st}&=&\bar m\ , \hspace{1.5cm} \av{ b_1\, b_2}_{st}=-\bar m\, 
\nn\\
\av{ a\da\, b_j}_{st}&=&\av{ a\, b_j}_{st}=\av{ b_j\, b_j}_{st}=0
\end{eqnarray}
where $\bar n$ and $\bar m$ are defined in Eq.~\rp{nm0}, 
that correspond to a squeezed state for the cavity field and a two-mode EPR entangled state for the mechanical resonators. We note that the effective number of thermal excitation for the Bogoliubov modes is zero, $\bar n_s=0$, when $\kappa_c'=0$, and it corresponds to $\bar m=\sqrt{\bar n(\bar n+1)}$ [see Eq.~\rp{nm0}]. In this case the steady state of the mechanical oscillators is a pure two-mode squeezed state. When $\kappa_c'\neq 0$, instead, the steady state is a thermal squeezed state also in the ideal limit of negligible mechanical noise (see Ref.~\cite{ZippilliPRA14} for similar considerations).  

In this case, the amount of entanglement of the mechanical oscillators, expressed in terms of the Logarithmic negativity, it is given by
\begin{eqnarray}\label{EN0}
E_N\al{ideal}&=&{\rm max}\pg{0,-\log\pt{2\bar n+1-2\bar m}}
\nn\\
&=&{\rm max}\pg{0,-\log\pt{S(0)}}
\end{eqnarray}
where $S(0)$ is the value of the squeezing spectrum of the driving field at its central frequency [see Eq.~\rp{S}]. This value is actually equal to the maximum entanglement between two spectral sidebands of the continuous-wave stationary squeezed field~\cite{ZippilliNJP15} which constitute our reservoir, as such it is the maximum entanglement which we expect one can extract by such a field.
Eq.~\rp{EN0} has been used to compute the solid thick grey lines reported in Figs.~\ref{fig_t}-\ref{fig_opt}.

\end{document}